# Atomic scale mapping of impurities in partially reduced hollow TiO$_2$ nanowires

*Joohyun Lim[1,‡,*], Se-Ho Kim[1,‡], Raquel Aymerich Armengol[1], Olga Kasian[1,2], Pyuck-Pa Choi[3], Leigh T. Stephenson[1], Baptiste Gault[1,4,*], & Christina Scheu[1]*

[1]Max-Planck-Institut für Eisenforschung GmbH, Max-Planck-Straße 1, 40237, Düsseldorf, Germany

[2]Helmholtz-Zentrum Berlin, Helmholtz-Institute Erlangen-Nürnberg, 14109 Berlin, Germany

[3]Department of Materials Science and Engineering, Korea Advanced Institute of Science and Technology (KAIST), 291 Daehak-ro, Yuseong-gu, Daejeon 34141, Republic of Korea

[4]Department of Materials, Royal School of Mines, Imperial College, London, SW7 2AZ, United Kingdom

## Abstract

The incorporation of impurities during the chemical synthesis of nanomaterials is usually uncontrolled and rarely reported because of the formidable challenge that constitutes measuring trace amounts of often light elements with sub nanometre spatial resolution. Yet these foreign elements influence functional properties, by e.g. doping. Here we demonstrate how the synthesis and partial reduction reaction on hollow TiO$_2$ nanowires leads to the introduction of parts-per-millions of boron, sodium, and nitrogen from the reduction reaction with sodium borohydride at the surface of the TiO$_2$ nanowire. This doping explains the presence of oxygen vacancies at the surface that enhance the activity. Our results obtained on model metal-oxide nanomaterials shed light on the general process leading to the uncontrolled incorporation of trace impurities that can have a dramatic effect on their potential use in energy-harvesting applications.

The morphology and porosity of functional nanomaterials determine a variety of properties such as efficiency, selectivity, and stability of (electro)catalysts for energy storage-, and sensor applications[5-7]. Another critical aspect is the presence of point defects such as vacancies or foreign elements that can affect the functionality of materials by modifying the band-gap, conductivity, catalytic property, or corrosion behavior[1-4]. Besides intentionally added dopants, trace impurities can be introduced within the material during synthesis. Their detection on a local scale remains a formidable challenge though and tackling it is critical to understand structure-property relationships and optimize strategies for the design of nanomaterials.

$TiO_2$ is one of the most widely studied metal oxide material because of its high abundance and superior performance as photocatalysts, catalyst supports, or electrode materials for solar-cells and metal-ion batteries[24,25]. Stoichiometric $TiO_2$ has a band gap of 3–3.2 eV depending on the crystal structure, which allows to absorb light only in the UV regime. Recently partially reduced $TiO_2$, also referred as black $TiO_2$, has been widely studied due to its visible light absorption enabled by a smaller band-gap and excellent metal-like conductivity[26,27]. These beneficial properties are generally attributed to the surface disorder associated with the presences of oxygen vacancies and titanium interstitials as well as other non-stoichiometric regions within the crystal inducing band gap states[49]. Controlling the presence of the metastable, defective and off-stoichiometric $TiO_2$ at the surface of nano-structured materials is critical to achieve enhanced properties.[N5N1]

Several approaches exist to synthesize black $TiO_2$ [ref]. For example, $NaBH_4$ can be used as a reductant under inert gas atmosphere, which has the advantage of a higher safety compared to using hydrogen gas for reduction.[28] Additionally, a lower reaction temperature can be used when $NaBH_4$ is involved in the reduction process, which makes this method applicable to $TiO_2$ grown on temperature sensitive transparent glass electrodes such as fluorine-doped tin oxide (FTO). $TiO_2$/FTO assemblies are typically used for photoelectrochemical- and solar-cell applications, and they would lose their good conductivity if exposed to higher temperature[29,30]. However, the presence and location of trace elements from the inert gas (e.g. nitrogen) or impurities from the reductant (sodium and boron) and their effects on the structure and properties of $TiO_2$ are still unclear[31]. Basically, all these elements can introduce disorder in the $TiO_2$ crystal lattice and can lead to a lower band-gap. N can act as an anionic dopant of $TiO_2$ since its atomic p-orbital levels are appropriate for narrowing the otherwise wide band-gap of $TiO_2$[38]. B in $TiO_2$ can have a synergy effect as O–Ti–B–N bonds narrow the bandgap and compensate the excess charge from N doping[50,51]. Interstitial B has been suggested to stabilize $Ti^{3+}$ species[N6]. The role of Na for the $TiO_2$ conductivity seems not clear yet, although Na can act as a recombination center or limiting the crystallization of $TiO_2$[47]. Besides being incorporated in the lattice, foreign elements from the synthesis can also be located at the very surface, either as clusters or individual atoms, which is in particular important for porous nanomaterials. Therefore, detecting and locating trace impurities with a sensitivity in range of parts-per-million as well as visualizing the distribution of oxygen vacancies and $Ti^{3+}$ defects in $TiO_2$ nanoporous materials is crucial to understand the influence of the synthesis route on the functional properties.

Complex three-dimensional (3D) nanomaterials are usually analyzed by transmission electron microscopy (TEM) enabling to assess their size, morphology, and crystal structure[N2]. Recent development in electron tomography, i.e. through acquisition of a tilt-series in TEM or scanning TEM (STEM), allows advanced 3D structural characterization[8]. Furthermore, the combination with energy dispersive X-ray spectroscopy (EDS) or electron energy loss spectroscopy (EELS) offers the opportunity for chemical analysis[9]. However, the distribution of surface disorder and oxygen vacancies has been barely analyzed in well-defined metal oxide nanostructures locally[N1,N7].

Atom probe tomography (APT) provides 3D elemental information with near-atomic resolution and a higher sensitivity than EDS and EELS, in particular for light elements[10,11]. Moreover, combined APT and high-resolution (S)TEM have recently enabled quantitative measurements of solute segregation at grain boundaries as a function of grain boundary character and misorientation[12,13]. Although APT has been mostly used to analyze bulk materials[14] and some semiconductor nanowires with dopants[15-18], it is still a burgeoning technique to investigate freestanding nanomaterials, in particular metal catalyst, metal oxide battery particles, or quantum dots[19-23]. Analyzing porous metal oxide nanomaterials using APT is highly challenging due to the low conductivity of metal oxides and the presence of pores. The latter renders the electrostatic field distribution very inhomogeneous and exacerbate reliable sample preparation technique. A step towards solving these issues could be the use of electrodeposition to fill the pores in microporous materials with a metal[20,32,33]

In the present work we studied partially reduced hollow $TiO_2$ 1D nanowires (R-HTNWs) using (S)TEM and APT as a model system to demonstrate the feasibility of our approach. The 3D morphology of the R-HTNWs was characterized using electron tomography while the local distribution of oxygen vacancies, more specifically the correlated $Ti^{3+}$ species, was analyzed using EELS in STEM mode. To enable APT analysis, electrodeposition was used to embed R-HTNWs in a Ni matrix, which also fills the hollow core of the NW. The 3D hollow structure of partially reduced $TiO_2$ was successfully reconstructed and the distribution of trace elements such as B, Na, and N, resulting from the reduction with $NaBH_4$ under $N_2$ atmosphere, were quantified. Our results show the ability to characterize the 3D structure and spatial distribution of impurities in porous metal oxide nanomaterials at an atomic resolution.

**Results**

**Characterization of R-HTNWs using EM** Hollow $TiO_2$ nanowires (HTNWs) were synthesized on a FTO glass substrate using a hydrothermal method and reduced to R-HTNWs using $NaBH_4$ as a reductant under $N_2$ flow. The details of the synthesis process are described in the Method section. Scanning electron microscopy (SEM) and TEM images show the hollow morphology of the R-HTNWs which are composed of an assembly of nanofingers (Figure 1a and 1b). This is confirmed by the side-view TEM image in Figure 1c. Lattice distances of 2.9 and 3.2 Å are obtained from one nanofinger of R-HTNWs, which are assigned to rutile (001) and (110), respectively (Figure 1d). Negative spherical aberration imaging in an aberration corrected TEM allows us to resolve atomic positions of Ti–O and Ti only along the rutile [110] viewing direction with different contrast (Figure 1e)[N3].

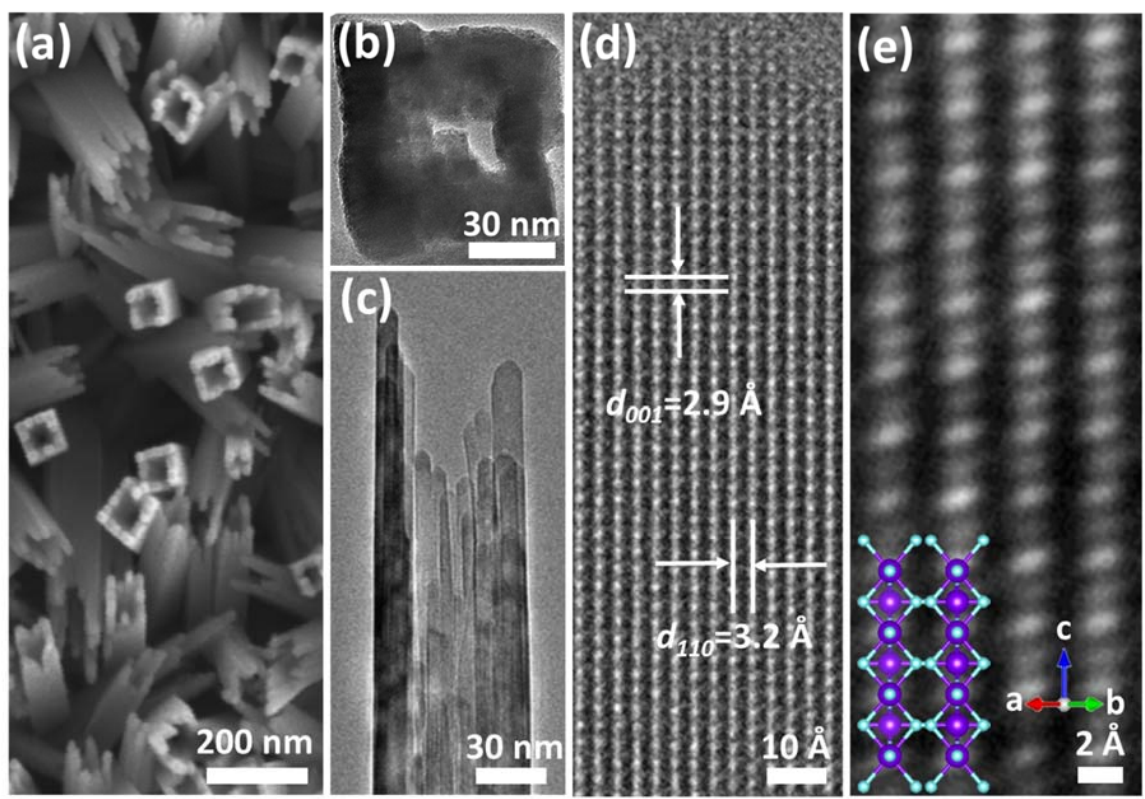

**Figure 1.** (a) SEM and (b-e) TEM images of R-HTNWs. (e) Atomic resolution image with the projected crystal structure of rutile $TiO_2$ in [110] viewing direction (purple spheres represent Ti and cyan spheres O atoms).

Chemical anlayese of R-HTNWs were performed with STEM-EDS and STEM-EELS. The side-view high-angle annular dark-field (HAADF)-STEM image and EDS elemental maps of Ti and O in Figure 2a show the hollow morphology with homogeneous distribution of both Ti and O. Similar information is obtained in top-view. Figure S2a presents STEM-EELS spectra of $Ti–L_{2,3}$ and O–K edges. The $Ti–L_{2,3}$ edge has four peaks, resulting from the spin-orbit splitting into $Ti–L_3$ ($2p_{3/2}$) and $Ti–L_2$ ($2p_{1/2}$) edges with further crystal field splitting of Ti 3d states into $t_{2g}$ and $e_g$. The four peaks here are characteristic for crystalline $TiO_2$ such as anatase and rutile[39,40]. The left shoulder of the $e_g$ peak of $Ti–L_3$ indicate the rutile nature of R-HTNWs[N1]. The O–K edge splitting indicates O 2p–Ti 3d hybridization states of $t_{2g}$ and $e_g$. Elemental maps were obtained using Ti–$L_{2,3}$ and O–K edges. Figure S2b presents annular dark-field (ADF)-STEM image and STEM-EELS maps for Ti–$L_{2,3}$ and O–K edges. The hollow morphology with nanofingers of the R-HTNWs is clearly visible in the EELS maps because of the thickness-sensitive signal. Figure S2c shows EELS data obtained from a single nanofinger of R-HTNW and HTNW (Figure S3d) to compare oxygen vacancies before and after the reduction reaction by $NaBH_4$. No peak splitting is observed in the Ti–$L_2$ and Ti–$L_3$ peaks of a R-HTNW single nanofinger and the peaks occur at slightly lower energy loss when compared to that of a HTNW single nanofinger. This result indicates the presence of oxygen vacancies with the partial reduction of $Ti^{4+}$ to $Ti^{3+}$, inducing the local structure distortion of the rutile structure[37]. For further understanding of the oxygen vacancy distribution of R-HTNW single nanofinger, EELS Ti–$L_{2,3}$ edges are obtained from one surface though the middle to the other surface with ~1 nm steps (Figure 2b). A high $I(L_2)_{Ti}$ to $I(L_3)_{Ti}$ indicates a high $Ti^{4+}$ concentration. The $I(L_2)_{Ti}/I(L_3)_{Ti}$ has the highest value in the core part of a single nanofinger and decreases toward both surfaces, revealing a surface specific introduction of oxygen vacancies and their gradient distribution. EELS Ti–$L_{2,3}$ edge from surface regions do not show the typical split into four peaks and a slightly lower energy loss. This indicates that the oxygen vacancies or other impurities stabilizing $Ti^{3+}$ are concentrated at the surface and the distorted rutile structure region.

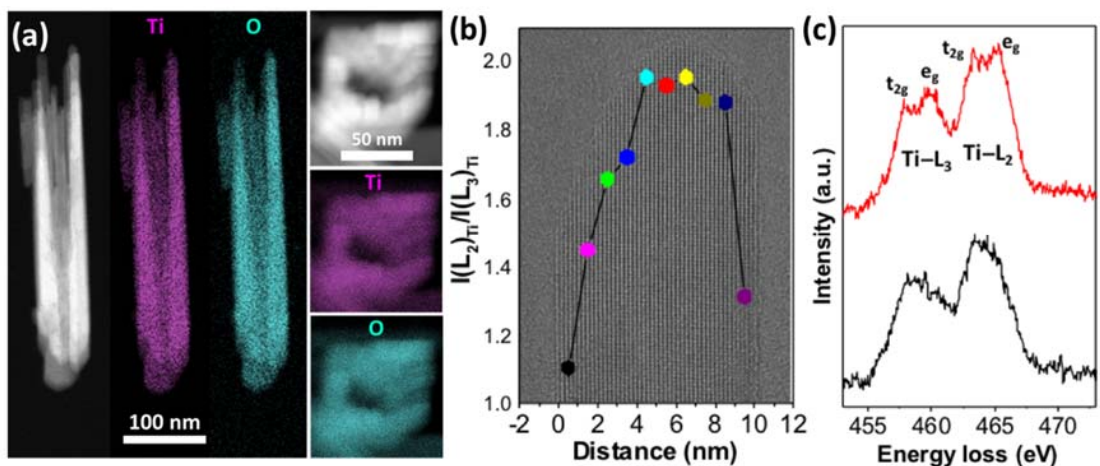

***Figure 2.*** *HAADF-STEM images and STEM-EDS maps for both side- and top-view of R-HTNWs. (b) TEM lattice image of a R-HTNW single nanofinger and overlaid $(L_2)_{Ti}/I(L_3)_{Ti}$ values determined from the EELS $Ti–L_{2,3}$ data at each position. The $I(L_2)_{Ti}/I(L_3)_{Ti}$ was obtained by averaging over an box with a size of $\sim 1 \times 10$ nm².with the long axis parallel to the wire surface. (c) $Ti–L_{2,3}$ edge for the region located at the surface (black spectrum) and $\sim 6$ nm away (red spectrum).*

To reveal the 3D morphology of an entire R-HTNWs, STEM tomography was performed. The data were also used to determine the regions of the APT reconstruction, where only a part of the wire is analyzed. The HAADF-STEM images from R-HTNWs tilted in the range of ±60º are reconstructed to a 3D tomogram. Figure 3a-c summarizes representative HAADF-STEM images taken at 0, −60, and +60 tilt angles, showing changing contrast by tilting the sample. The 3D reconstructed volume shows the nanofingers as well as the empty space inside the nanowire (Figure 3d). The cross-sectional cut through the reconstructed 3D volume presents the hollow nature of R-HTNWs in detail (Figure 3e). EM successfully provides the 3D morphology, chemical composition, and crystal structure of R-HTNWs, which are of importance for various functional properties of porous nanomaterials. However, the amount of possible trace impurities such as B, Na, and N in the R-HTNWs stemming from the synthesis procedure are too low to be detectable by EDS or EELS and their presence and location remain unknown. As such impurities can dramatically change the local crystal structure and properties of TiO$_2$ as described above, APT measurements are used to track their position.

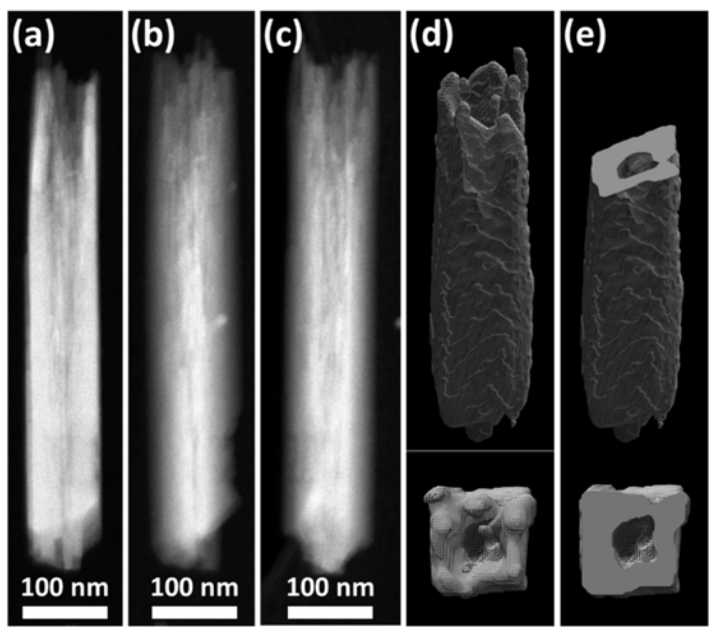

**Figure 3.** Representative HAADF-STEM images of a R-HTNW from (a) 0, (b) −60, and (c) +60 tilt angles. (d) 3D reconstructed volume and (e) sections of the R-HTNW.

## Characterization of R-HTNWs using APT

The low electrical conductivity of TiO$_2$ and the presences of pores in the R-HTNWs were obstacles for conventional APT characterization. Filling pores with a highly conductive material has been shown to be a solution for stable and homogeneous electrostatic field distribution[20,32,33], but has not been demonstrated for poorly conducting, porous nanomaterials. Here, we co-electrodeposited R-HTNWs, detached from the FTO along with

Ni to form a film on a Si substrate, which allowed to prepare APT specimen using focused ion beam (FIB) milling[N4]. Metal oxide nanoparticles (i. e. $Al_2O_3$, $TiO_2$, and $La_2O_3$) have been reported as inert particles during co-electrodeposition process of Ni films[41-43]. Therefore, this single-step method, without the electrophoresis step, is simpler than the previously reported two-step electrodeposition method[20]. The electrodeposition depends strongly on the surface heterogeneity[44,45]. In our case, we observed protrusions on the co-electrodeposited Ni surface, which was indicative of the presence of the embedded nanowires (see Figure S3). The presence of the nanowires inside of the protrusion is confirmed by cross-sectioning one of these regions with the FIB (see Figure S4a). The hollow and 1D nanowire morphology of the R-HTNWs are preserved after co-electrodeposition with Ni. The pores in the R-HTNWs are successfully filled with Ni without any voids, which could induce inhomogeneous field evaporation during APT measurements. Every protrusion we investigated contained R-HTNWs helping us to guide the fabrication of site-specific APT specimen. One of these regions was then lifted out, deposited onto a support, sliced to only contain the R-HTNWs, and sharpened into a needle-like specimen (see Figure S4b and 4c).

Details of the APT analysis of the R-HTNWs embedded in Ni can be seen in Figure S5 and Figure S6. The major peaks in the mass-to-charge-ratio spectrum can be assigned to Ti–O molecular ions and the electrodeposited Ni matrix in single and double charged states (see Table S1). A strong $O^+$ peak originating from the R-HTNWs is detected at 16 Dalton (Da). Several peaks can be clearly assigned to atomic species introduced during the reduction of R-HTNWs. The peaks at 14 and 23 Da are $N^+$ and $Na^+$ ions, respectively, and the peaks at 10 and 11 Da correspond to $B^+$ ions. N can be incorporated during the heating process in presence of a nitrogen flow whereas B and Na can result from the thermal degradation of $NaBH_4$ during the reduction process.

A reconstructed 3D atom map containing a part of the R-HTWN embedded in Ni is shown in Figure 4a and Figure 4b for the hollow nanowire and single nanofingers regions, respectively. Reconstructed Ti (purple) and O atoms (cyan), and an iso-concentration surface of 15 at.% for Ti (purple) are presented within the Ni matrix (yellow). The 20 nm and 10 nm thin slices for the reconstructed hollow nanowire region and a single nanofinger region, respectively, viewed along y-axis show the distribution of other elements such as B, Na, and N. These elements are concentrated along the surface of R-HTNWs and the surface centered distribution of impurities is in accordance with that of oxygen vacancies from EELS.

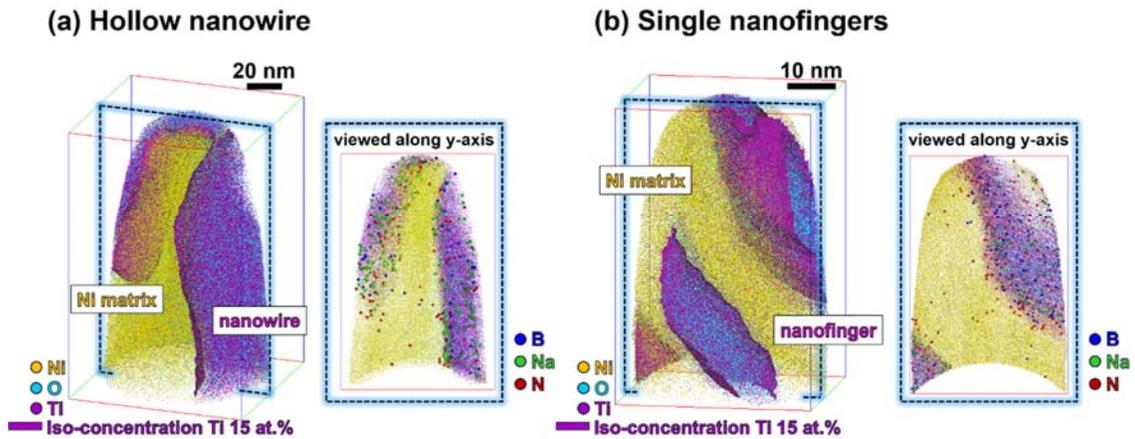

**Figure 4.** 3D atom map of (a) hollow nanowire and (b) single nanofingers of R-HTNWs embedded within Ni using an iso-concentration surface of Ti at 15 at.% (purple) with the thin slice viewed along y- direction in R-HTNW (20 and 10 nm for the hollow nanowire and single nanofingers region, respectively). Yellow, cyan, purple, blue, green, and red dots represent the reconstructed atomic position of Ni, O, Ti, B, Na, and N, respectively.

The distribution of B, Na, and N atoms in the R-HTNWs is quantified by a composition profile calculated as a function of the distance to the isosurface (proximity histograms) defined above and plotted in Figure 5a. The main elements concentration of Ti and O reach 34 and 63 at.%, respectively, in the bulk of the nanowire region. The ratio of Ti to O concentration is ~1.9 which is close to the stoichiometry of $TiO_2$. Minor elements such as B, Na, and N are detected inside this bulk region too, which indicates that these elements are incorporated during the synthesis procedure of the R-HTNWs. In this part of the nanowire, the atomic concentrations of B, Na, and N are 0.31, 0.22, and 0.026 at.%, respectively. The surface of the R-HTNW, marked by the intersection between the Ni-matrix and Ti–O, see a strong enrichment in B, Na, and N.

The interfacial excess value ($\Gamma_i$) of minor elements (B, Na, and N) from both hollow nanowire and single nanofingers regions can be determined form the impurity concentration detected at the surface (see Supporting text for calculation details). The values of $\Gamma_i$ after background correction, for both the hollow nanowire and single nanofinger, are presented in Table S2. They differ by a factor of ~2. This is because for the calculation all four {110} surface planes are considered as the exposed surface area, two of which are actually blocked by other nanofingers (Figure S8).

To address the origin of these minor elements further, an APT analysis was conducted on the as-grown and non-reduced HTNWs. Figure S7 shows the atom map and element composition profile of such a HTNW. Boron is not detected (see Table S2), confirming that the source of boron impurity is the $NaBH_4$ used for reduction. A low amount of Na is detected in the HTNWs, indicating that the presences of Na in R-HTNWs can be from the $NaBH_4$ used for the reduction as well as from the diffusion of Na from the glass-based FTO substrate during the hydrothermal synthesis of the HTNWs[46,47]. The low amount of N detected can originate from the nitrogen atmosphere during reduction or from adsorption during specimen transfer or their processing. A proximity histogram from a single nanofinger region in R-HTNW shows similar result to the hollow nanowire region (Figure 5b). The result reveals that impurities of Na, B, and N have a strong tendency to be deposited on the exposed {110} surface of the R-HTNWs during synthesis or the reduction.

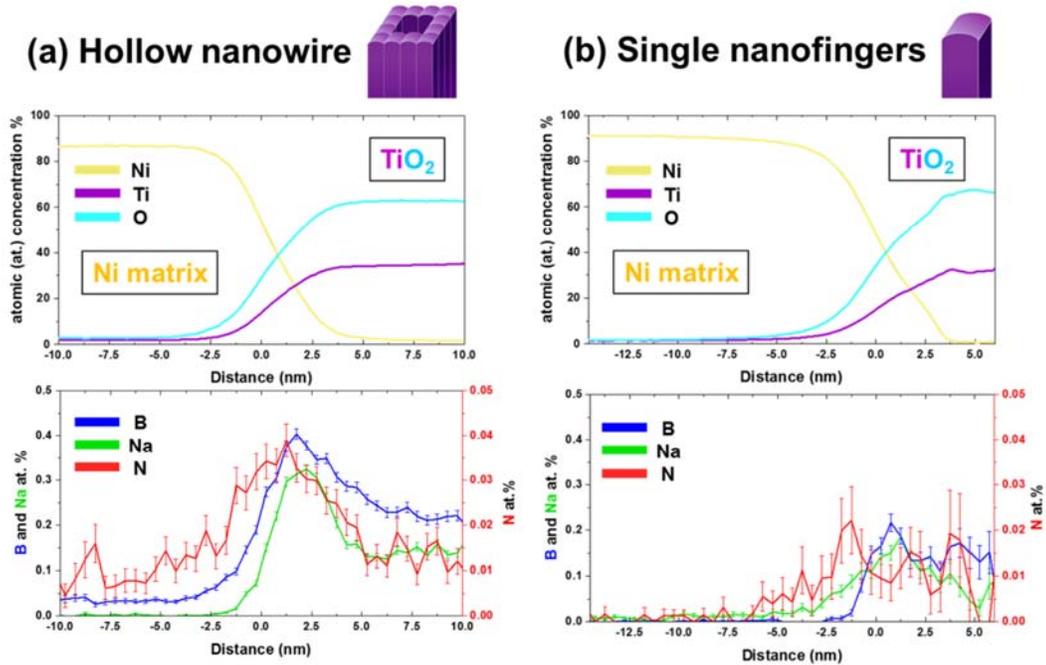

**Figure 5.** *Proxigram concentration profile of (a) hollow nanowire and (b) single nanofingers region in R-HTNWs. (top) Major (Ni, Ti, O) and (bottom) minor (B, Na, N) elements along with iso-concentration of 15 at.% for Ti.*

We have successfully resolved the 3D morphology of R-HTNWs using EM and APT. The distribution of oxygen vacancies on single nanofinger was characterized using EELS and trace impurities of B, Na, and N were mapped in atomic scale using APT. We could unequivocally demonstrate that the impurities originate from $NaBH_4$ that was used as a reductant for $TiO_2$. This uncontrolled ingress of impurities was revealed by APT, as enabled by an advanced sample preparation technique whereby these hollow nanomaterials were embedded within a metallic matrix. Both oxygen vacancies and these impurities were concentrated at the surface of the R-HTNW.

These surface defects need to be considered when discussing structure - property relationships of R-HTNWs as their impact can be very large [N9]. First, the surface impurities, which are unintentionally added to the system, contribute to distorting the rutile crystal struture. This distortion is responsible for higher carrier mobility leading to metallic-like conductivity[N10]. The surface impurities also stabilize $Ti^{3+}$ on the surface as interstitial B allows to have sufficient $Ti^{3+}$ species in $TiO_2$ (B-O-$Ti^{3+}$)[N6N8]. This is supported by out STEM-EELS measurements. B and N synergistically reduce the band-gap of $TiO_2$, increasing the performance for photo(electro)catalyst and energy storage applications[50,51]. Additionally, surface B increases electrocatalytic properties in fuel cell applications due to the high oxygen adsorption ability[N8]. All these examples show that the detection of trace elements enabled by the combination of advanced electron microscopy and atom probe tomography is necessary to further understand the functional properties of these and other nanomaterials. The lack of precise, quantitative characterization of impurities likely explain contradicting results reported in the literature, since different amounts of trace elements might be introduced during the synthesis, and were, up to now, most often not considered. The approach discussed herein could readily be deployed to a vast array of different materials systems and hence lead to a better control over the integration of impurities with an aim to enhance the materials' functional properties.

**Methods**

**Synthesis of R-HTNWs** Rutile TiO$_2$ nanowires were grown on a FTO glass substrate using a modified hydrothermal method[34,35]. They are composed of finer nanofinger bundles surrounded by compact and bigger outer nanofingers (Figure S1a and S1b). HTNWs were obtained by selective core etching of TiO$_2$ nanowires with a 1:1 (v/v) mixture solution of deionized water and hydrochloric acid (HCl, 36 wt%, Sigma Aldrich) in an autoclave reactor at 180ºC for 2h. The porous and finer inner nanofinger bundles could selectively be etched out in HCl solution because of their higher Ti–OH surface area where Cl$^-$ can easily attack (Figure S1c), resulting in hollow morphologies[36,37]. Finally, R-HTNWs were prepared by the reduction of HTNWs with sodium borohydride (NaBH$_4$, Sigma Aldrich) under N$_2$ flow at 400 ºC for 2h with a heating rate of 10 ºC/min. The obtained R-HTNWs were washed using deionized water and dried at room temperature.

**Co-deposition of R-HTNWs within an electrodeposited Ni** As-synthesized R-HTNWs were encapsulated within a Ni film according to the modified procedure developed by Kim et al[20]. For Ni electrolyte preparation, 15 g of nickel sulfate hexahydrate (NiSO$_4$-6H$_2$O, Sigma Aldrich) and 2 g of boric acid (H$_3$BO$_3$, Sigma Aldrich) were dissolved in 50 mL of distilled water (18.2 MΩ-cm) from the AQUA Solution type 1. Then, as-synthesized R-HTNWs were dispersed in the Ni electrolyte using a sonicator for 10 min and the complex solution was poured into a vertical cell for co-electrodeposition process. Co-deposition was carried out in a specially designed vertical cell including a Cu substrate and a Pt-mesh counter electrode as shown in Figure S9a and S9b. In order to homogenously deposit Ni on the Cu substrate and effectively deposit the nanowires within the Ni layer, a vertical cell is designed in a conical shape having that the surface area of Pt counter electrode (2 cm$^2$) is larger than that of the Cu working electrode (0.2 cm$^2$). The one-step co-electrodeposition was performed at a constant current of 19 mA for 500s. In Figure S9c, a photograph of the co-electroplated Ni and R-HTNWs on a Cu substrate is shown.

**Electron microscopy characterization** SEM was performed to investigate the morphology of the R-HTNWs (Gemini 500, Zeiss, in-lens detector, 2 kV). (S)TEM was performed using a FEI 60-300 Titan Themis operated at 300 kV with a Cs-corrector for the probe forming lens. The chemical composition of R-HTNWs was analyzed by EDS in the STEM mode. EELS data were acquired in the STEM mode using a dual channel acquisition mode[55] with a dispersion of 0.05 eV per channel and a pixel time of 2-3s using a Gatan software. All spectra were corrected for channel to channel gain variations and dark current[56]. The background was subtracted using a standard power law. Post-edge background of Ti–L$_{2,3}$ edge EELS were further removed with a double arctangent step function.

**Electron tomography** HAADF-STEM images were acquired from the R-HTNWs every 5º tilt angle over ±60º tilting range for the 3D reconstruction using a JEOL JEM-2200FS TEM at 200 kV. The obtained STEM images were aligned using the TomoJ plugin installed in ImageJ. Simultaneous iterative reconstruction (SIRT)[57-59] and discrete algebraic reconstruction technique (DART)[60,61] were applied to reconstruct the final 3D volume of the R-HTNWs.

**APT characterization** A needle-shape specimen was prepared from the co-deposited sample using FIB (Helios NANOLAB 600i, FEI) for APT measurement to identify the chemical composition and position for each element[62]. APT analyses were performed using a local electrode atom probe (Cameca LEAP 5000 XS system) in pulsed UV laser mode at a detection rate of 1 %, a laser pulse energy of 80 pJ, and a pulse frequency of 125 kHz. The specimen temperature was set to ~50 K during analysis. The set parameter was based on the recent report on natural rutile ($TiO_2$) measurement for atom probe tomography[63]. Data reconstruction and analyses were performed using the commercial software Imago visualization and analysis system standard (IVAS) 3.8.2 developed by Cameca Instruments. All 3D atom maps presented in this paper were reconstructed using the standard voltage reconstruction protocol.

**Corresponding Author**

Correspondence and requests for materials should be addressed to J.L. (email: j.lim@mpie.de) or to B.G. (email: b.gault@mpie.de) or C.S. (email: c.scheu@mpie.de)

**Author Contributions**

‡ J.L. and S.-H. K. contributed equally. J.L. performed the synthesis and the (S)TEM work (HAADF imaging, EELS, EDS, tomography). R. A. A. contributed to the synthesis. S.-H. K. performed the co-electrodeposition, atom probe specimen preparation and analysis with support from L.T.S. and B.G.. O. K. and P.-P. C. supported on electroplating and co-electrodeposition process. J.L., S.-H. K., C.S. and B.G. designed overall experiment and drafted the manuscript. All authors then contributed and have given approval to the final version of the manuscript.

**Acknowledgements**

J. L. acknowledges for the financial support from Alexander von Humboldt Foundation. S.-H.K., L.T.S., and B.G. acknowledge financial support from the ERC-CoG-SHINE-771602. We thank A. Folger for helping with the synthesis.

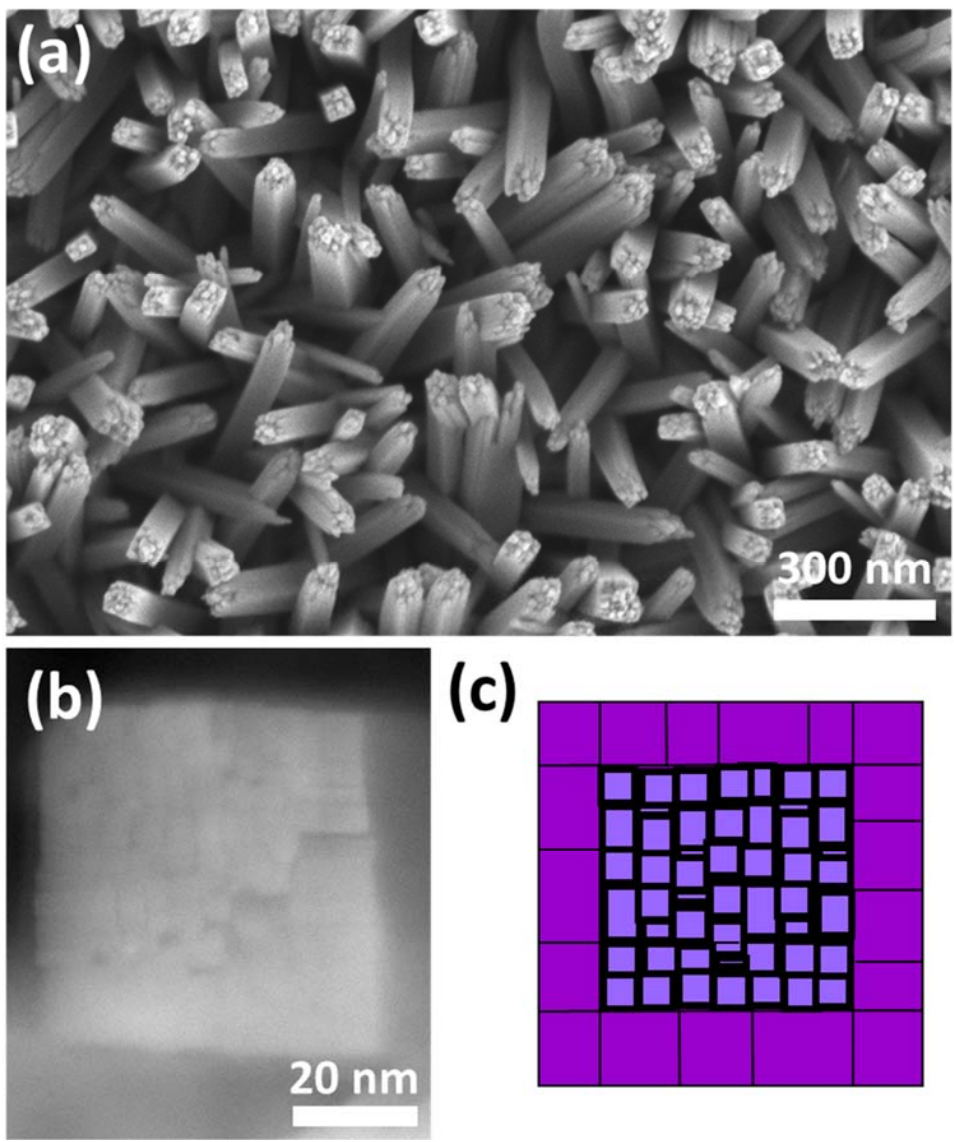

**Figure S1.** Top view of (a) SEM, (b) STEM, and (c) schematic images of as-synthesized TiO$_2$ 1D nanowires.

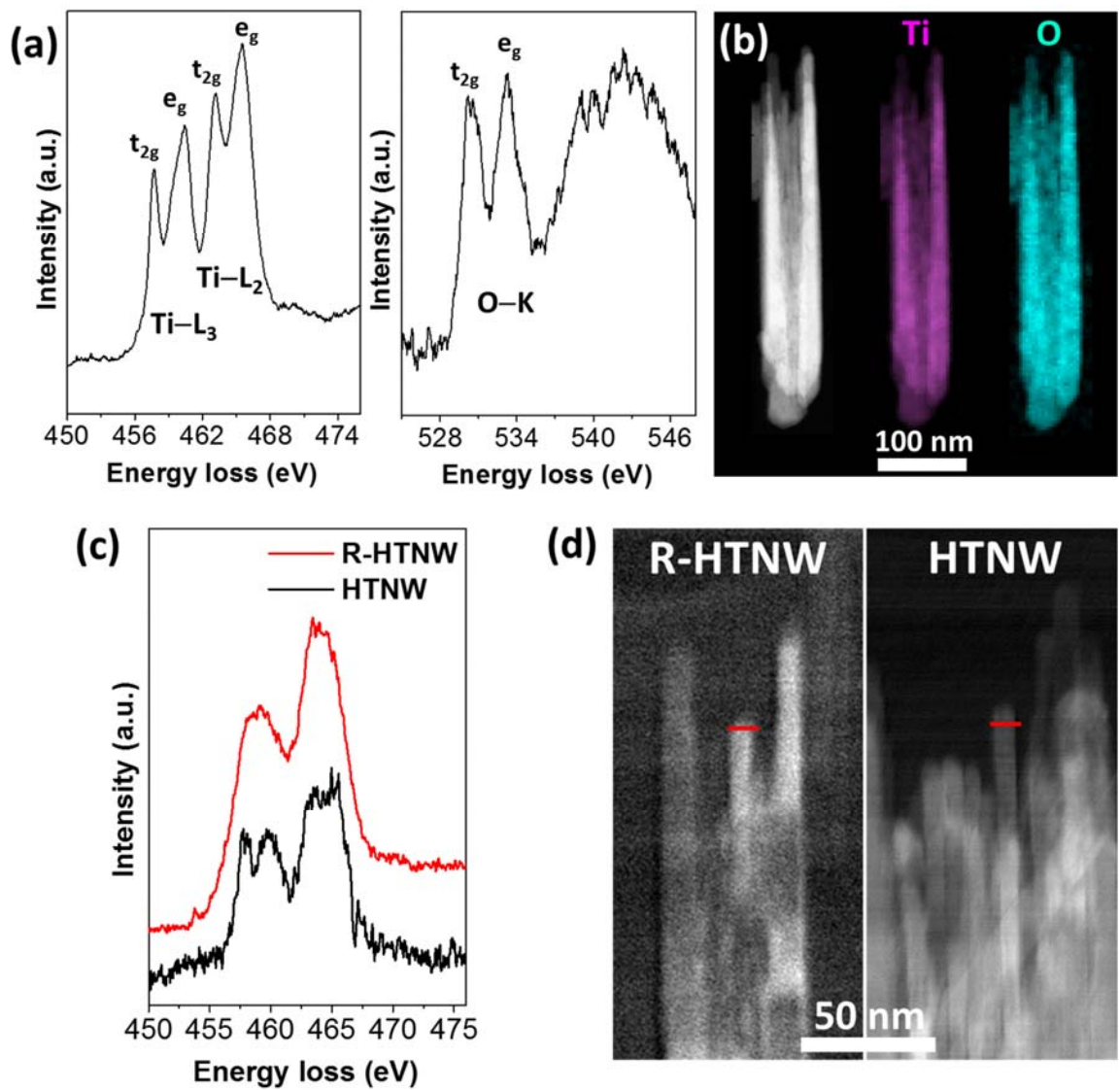

**Figure S2.** (a) STEM EELS Ti–L$_{2,3}$ and O–K edges of R-HTNW. (b) ADF-STEM image and STEM-EELS maps of R-HTNW. (c) EELS Ti–L$_{2,3}$ edge by line-scanning for single nanofinger of R-HTNW and HTNW. (d) STEM images of nanofingers where EELS line-scans were obtained (red line).

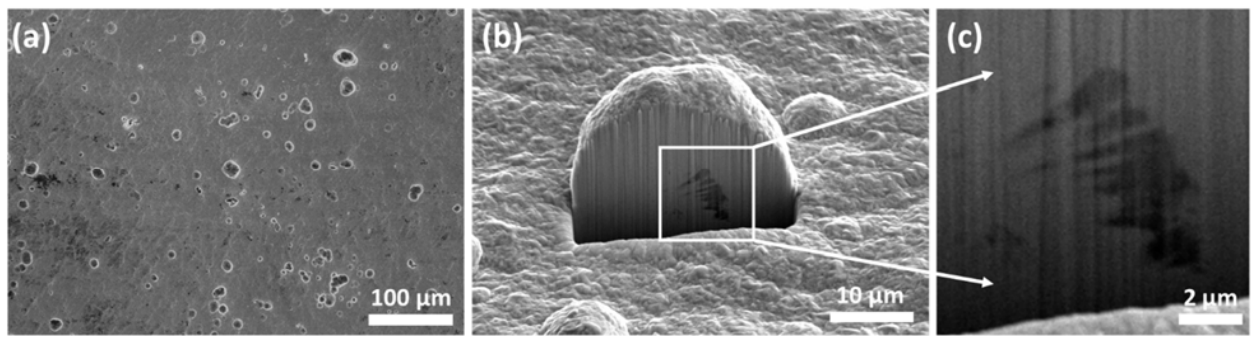

**Figure S3.** SEM images of (a) protrusions on the surface of Ni co-deposited with R-HTNWs and (b) and (c) cross-section of a protrusion.

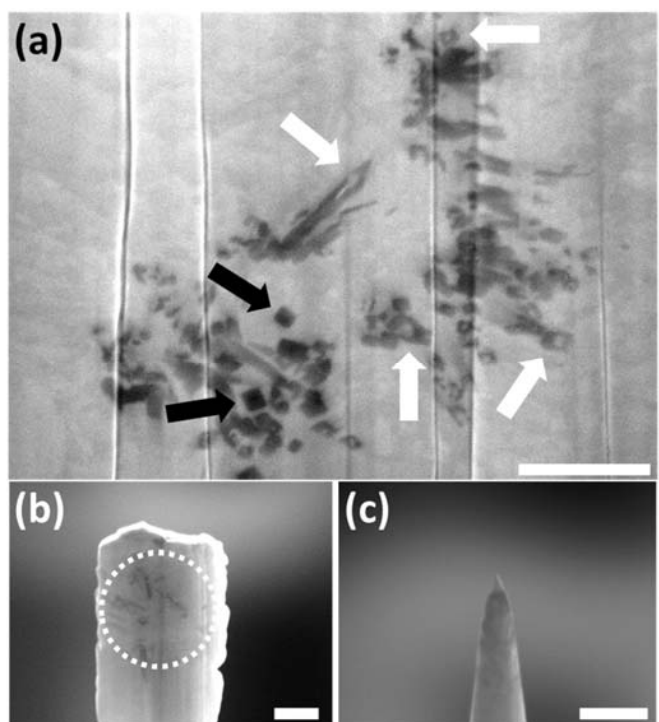

**Figure S4.** (a) Cross-sectional SEM image of R-HTNWs embedded in Ni. SEM images of the cut and lifted out area (b) before and (c) after final sharpening process. White scale bars indicate 500 nm.

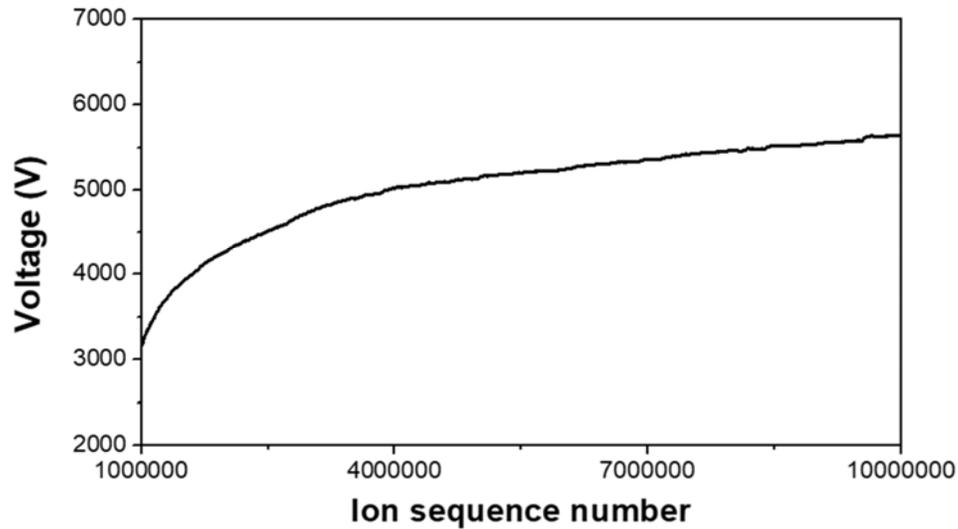

**Figure S5.** Voltage history curve for R-TNWS in Ni matrix during evaporation. The steady increase in voltage over the course of the APT measurement indicates a high reliability of our specimen preparation approach and the complete filling of the empty space in R-TNWS.

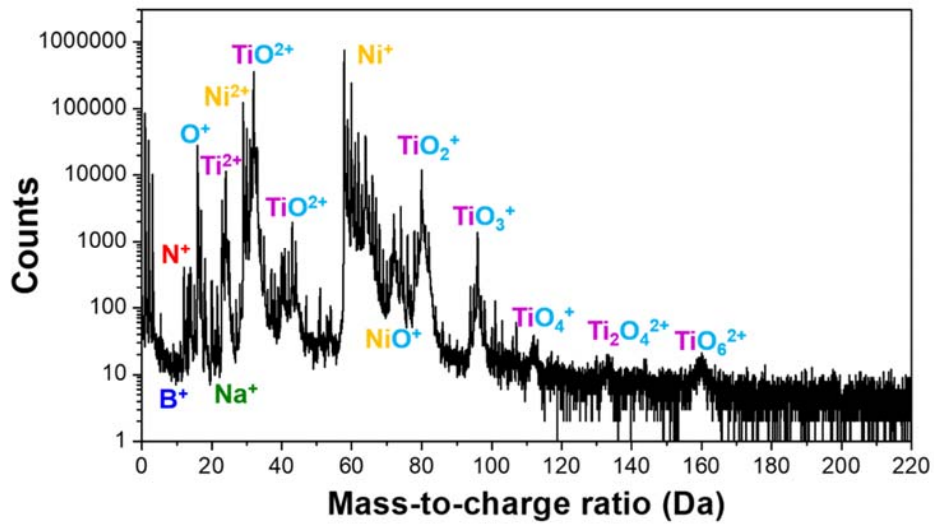

**Figure S6.** Overall mass spectrum of R-HTNWs embedded in Ni matrix.

**Table S1**. List of analyzed peak numbers from R-HTNWs embedded in Ni matrix.

| Chemical species | Charge state | Mass-to-charge ratio, m/z (Da) |
|---|---|---|
| Ni | 1+ | 58, 60, 61, 62, 64 |
| | 2+ | 29, 30, 30.5, 31, 32 |
| NiH | 1+ | 59, 61, 62, 63, 65 |
| NiO | 1+ | 74, 75, 76, 77, 78, 79, 80, 81 |
| | 2+ | 37, 37.5, 38, 38.5, 39, 39.5, 40, 40.5 |
| Ti | 1+ | 56, 57, 58, 59, 60 |
| | 2+ | 23, 23.5, 24, 24.5, 25 |
| TiO | 1+ | 62, 63, 64, 65, 66, 67 |
| | 2+ | 31, 31.5, 32, 32.5, 33, 33.5 |
| TiO$_2$ | 1+ | 78, 79, 80, 81, 82, 83, 84 |
| | 2+ | 39, 39.5, 40, 40.5, 41, 41.5, 42 |
| TiO$_3$ | 1+ | 94, 95, 96, 97, 98, 99, 100, 101 |
| | 2+ | 47, 47.5, 48, 48.5, 49, 49.5, 50, 50.5 |

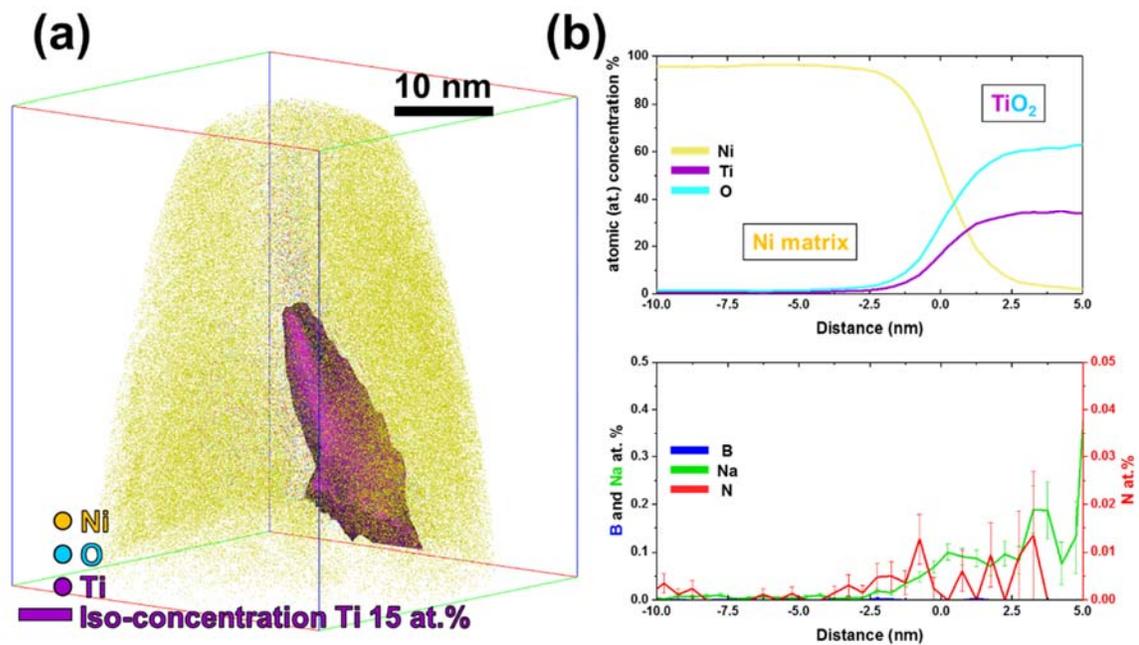

**Figure S7**. (a) 3D atom map of a HTNW embedded within Ni using an iso-concentration surface of Ti at 15 at.% (purple). (b) Proxigram concentration profile of major (Ni, Ti, O) and minor (B, Na, N) elements along iso-concentration of 15 at.% Ti. Yellow, cyan, purple, blue, green, and red dots represent the reconstructed atomic position of Ni, O, Ti, B, Na, and N, respectively.

**Table S2**. Bulk composition for for B, Na, and N in R-HTNWs and HTNW

| Elements (at.%) | Hollow nanowire (R-HTNW) | Single nanofingers (R-HTNW) | HTNW |
|---|---|---|---|
| B | 0.31 | 0.35 | - |
| Na | 0.22 | 0.19 | 0.21 |
| N | 0.026 | 0.028 | 0.013 |

**Table S3**. Surface density values for B, Na, and N for hollow nanowire and single nanofinger.

| $\Gamma_i$ (atom/nm$_2$) | Hollow nanowire (R-HTNW) | Single nanofingers (R-HTNW) | Ratio |
|---|---|---|---|
| B | 0.87 | 0.52 | 1.7 |
| Na | 0.49 | 0.26 | 1.9 |
| N | 0.086 | 0.047 | 1.8 |

**Supplementary Text**

<u>Surface density of impurities</u>

The Gibbsian interfacial excess ($\Gamma_i$) of minor elements such as B, Na, and N atoms at the R-HTNWs/Ni matrix interface are calculated using:[43]

$$\Gamma_i = \frac{1}{A\mu} N_i^{excess} = \frac{1}{A\mu} \left( N_i^{total} - N_i^{Ti} - N_i^{Ni} \right)$$

Here $A$ is the surface area from the R-HTNWs and $\mu$ is the detection efficiency (~80%) of the LEAP 5000 XS [44]. The excess number of impurity element $i$ associated with the TiO$_2$/Ni interface ($N_i^{excess}$) is determined by comparing total number of $i$ element ($N_i^{total}$) to a reference system where the number of element $i$ in R-HTNWs ($N_i^{Ti}$) and Ni matrix ($N_i^{Ni}$) is used.

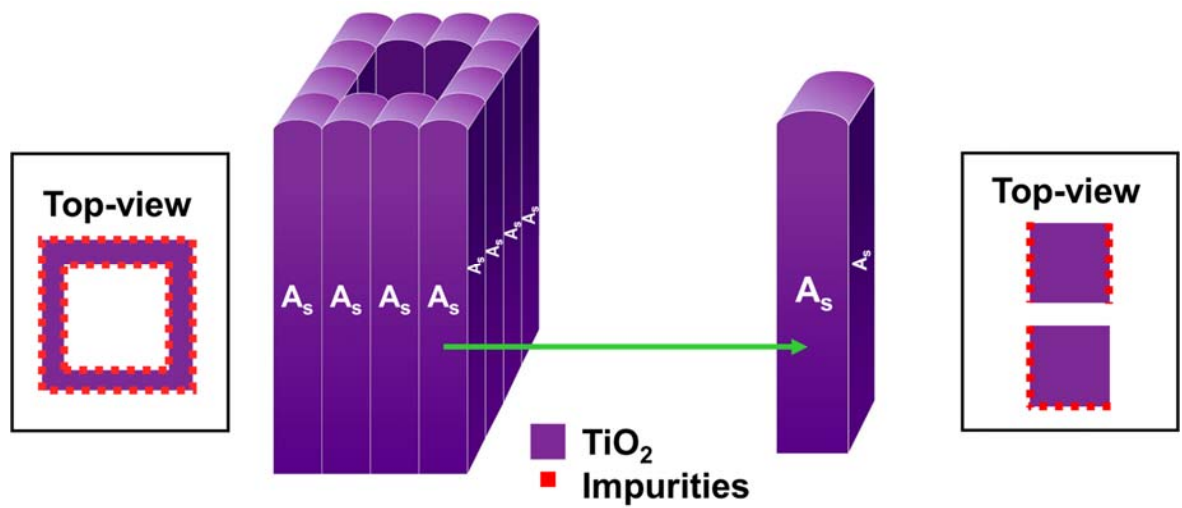

**Figure S8**. Schematic image for exposed surface of hollow nanowire and single nanofinger where impurities mainly present. Inset image of top-view of a single nanofinger demonstrates (top) side and (bottom) edge nanofinger from assembly with possible impurities distribution.

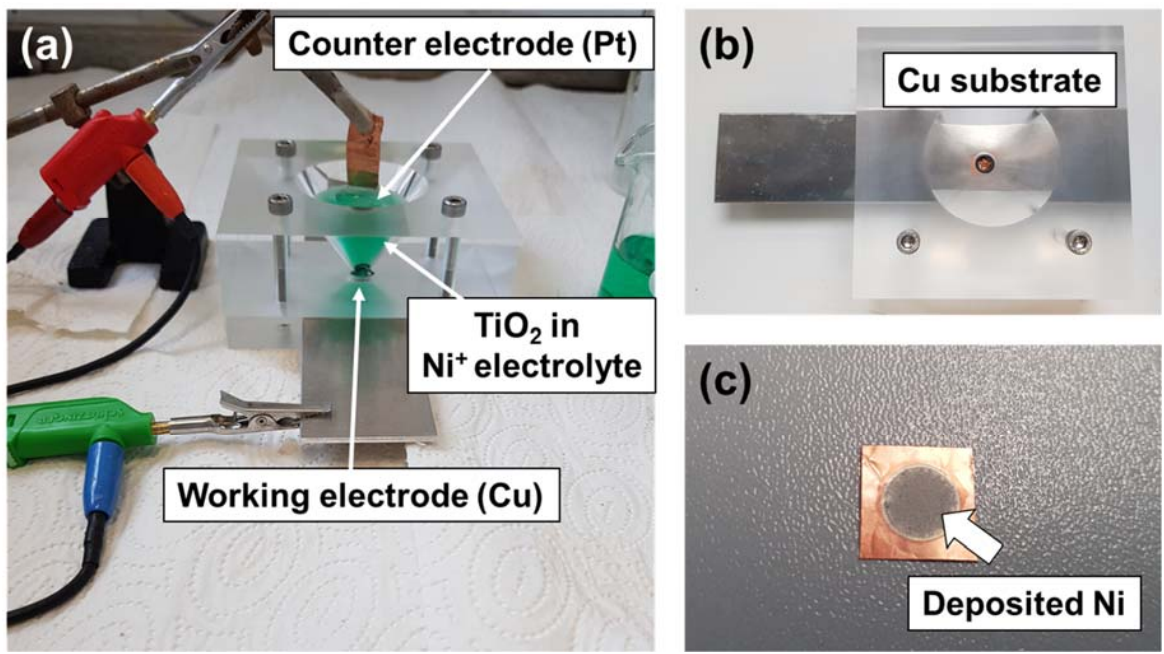

**Figure S9.** Photographs of the one-step co-deposition set-up. (a) A designed vertical cell with a Pt counter electrode (top) and a Cu working electrode (bottom). The $TiO_2$ dispersed Ni electrolyte is poured into the cell for co-deposition process. (b) Top view of the designed cell. (c) An example of the co-deposited sample of the R-HTNWs and Ni film.